\documentclass[aps,showpacs,prd,twocolumn]{revtex4}
\usepackage{graphicx}
\usepackage{amssymb}
\usepackage{amsmath}
\usepackage{color}
\usepackage{float}

\begin{document}

\title{Generalized self-dual Maxwell-Chern-Simons-Higgs model}
\author{D. Bazeia$^{1,2}$, R. Casana$^{3}$, E. da Hora$^{1,4}$ and R. Menezes%
$^{2,5}$.}
\affiliation{$^{1}${Departamento de F\'{\i}sica, Universidade Federal da Para\'{\i}ba,
58051-970, Jo\~{a}o Pessoa, Para\'{\i}ba, Brazil.}\\
$^{2}${Departamento de F\'{\i}sica, Universidade Federal de Campina Grande,
58109-970, Campina Grande, Para\'{\i}ba, Brazil.}\\
$^{3}${Departamento de F\'{\i}sica, Universidade Federal do Maranh\~{a}o,
65085-580, S\~{a}o Lu\'{\i}s, Maranh\~{a}o, Brazil.}\\
$^{4}${Grupo de F\'{\i}sica Te\'{o}rica Jayme Tiomno, S\~{a}o Lu\'{\i}s,
Maranh\~{a}o, Brazil.}\\
$^{5}${Departamento de Ci\^{e}ncias Exatas, Universidade Federal da Para%
\'{\i}ba, 58297-000, Rio Tinto, Para\'{\i}ba, Brazil.}}

\begin{abstract}
{We present a consistent BPS framework for a generalized
Maxwell-Chern-Simons-Higgs model. The overall model, including its self-dual
potential, depends on three different functions, $h\left( \left\vert \phi
\right\vert ,N\right) $, $w\left( \left\vert \phi \right\vert \right) $ and $%
G\left( \left\vert \phi \right\vert \right) $, which are functions of the
scalar fields only. The BPS energy is proportional to the magnetic flux when 
$w\left( \left\vert \phi \right\vert \right) $ and $G\left( \left\vert \phi
\right\vert \right) $ are related to each other by a differential
constraint. We present an explicit non-standard model and its topologically
non-trivial static configurations, which are described by the usual radially
symmetric profile. Finally, we note that the non-standard results behave in
a similar way as their standard counterparts, as expected, reinforcing the
consistence of the overall construction.}
\end{abstract}

\maketitle

\section{Introduction}

Topologically non-trivial structures have been intensively studied in many
areas of physics \cite{n5}. In particular, in the context of classical field
theories, these structures are described as finite-energy solutions to some
non-linear models. In this case, such models must be endowed by a
spontaneous symmetry breaking potential for the matter self-interaction,
since topological solutions are formed during symmetry breaking phase
transitions.

In this context, the most common topological defects are kinks \cite{n0},
vortices \cite{n1} and magnetic monopoles \cite{n3}: the first ones are
one-dimensional structures described by a single real scalar field, while
vortices and monopoles are two- and three-dimensional configurations arising
as static solutions of some Abelian- and a non-Abelian-Higgs models.

In particular, vortices are stable configurations described by radially
symmetric profile. The simplest version of such structures arises as
electrically non-charged solution of a planar Maxwell-Higgs model endowed by
a fourth-order Higgs potential. However, under special {circumstances},
vortices also arise as electrically charged configurations of some
Chern-Simons- and Maxwell-Chern-Simons-Higgs (MCS-Higgs) models \cite{mcsh}.
In all these cases, such structures can be found as numerical solutions to a
set of first-order differential equations, named
Bogomol'nyi-Prasad-Sommerfield (BPS) ones \cite{Bogo}. \ In this case, as
finite-energy solutions, they have interesting applications, mainly
concerning the superconductivity phenomena \cite{n1}.

Moreover, during the last years, beyond the standard models cited above,
non-usual ones have been intensively studied. These theories, generically
named \textit{k-field models}, are endowed by non-usual kinetic terms, which
change the dynamics of the overall model in a non-standard way. Here, it is
important to point out that the motivation regarding such generalization
arises in a rather natural way, in the context of string theories.

In fact, k-field theories have been used as effective models mainly in
Cosmology, as an attempt to explain the actual accelerated inflationary
phase of the universe \cite{n8} via the so-called k-essence models \cite{n7h}%
. Furthermore, they have been applied in the study of strong gravitational
waves \cite{n13}, dark- \cite{n11} and tachyon-matter \cite{n10}, and others 
\cite{n14}.

In such a non-standard scenario, topologically non-trivial configurations,
named \textit{topological k-solutions}, can exist even in the absence of a
symmetry breaking potential \cite{n14a}, from which one notes that the
existence of such structures are quite sensible to the presence of
non-standard kinetic terms. On the other hand, as an attempt to study
topological k-solutions via the comparison between them and their canonical
counterparts, some of us have already considered the existence of such
solutions in the context of symmetry breaking k-field models \cite{n14b}.
Moreover, interesting results concerning these models and the corresponding
solutions can be found in Ref.\cite{n14c}.

In general, given non-trivial kinetic terms, k-field models can be highly
non-linear, and the corresponding k-solutions can be quite hard to find. In
this case, the development of a consistent self-dual theoretical framework
is quite useful and desirable, since it helps to find topological
k-structures as solutions to some non-standard BPS equations. Here, as in
the usual approach, such equations can be obtained via the minimization of
the energy functional related to the non-standard model, the resulting
self-dual k-solutions being the minimal energy configurations possible ever.

In this sense, in a recent work, some of us have presented a BPS theoretical
framework consistent with a generalized self-dual Maxwell-Higgs model
endowed by non-usual kinetic terms to both gauge and scalar fields \cite%
{n14d}. Now, we introduce an extension of that work. Here, we develop a
general first-order approach consistent with a non-standard self-dual
MCS-Higgs model. The overall model, including its self-dual potential,
depends on three different functions, $h\left( \left\vert \phi \right\vert
,N\right) $, $w\left( \left\vert \phi \right\vert \right) $ and $G\left(
\left\vert \phi \right\vert \right) $, which are functions of the scalar
sector of the model. In this context, we look for topologically non-trivial
configurations via the usual radially symmetric static \textit{Ansatz}, and
the finite-energy generalized numerical solutions we found behave in the
same general way as their standard counterparts, as expected.

This letter is outlined as follows: in the next Sec.\ref{general}, we
introduce the generalized model and develop the BPS\ framework which allows
us to get to its non-standard BPS equations. Also, we verify the consistence
of the overall construction by using it to develop a generalization of the
usual MCS-Higgs case, the general model being controlled by two real
parameters, $\alpha $ and $b$. Then, in Section \ref{numerical2}, we perform
the numerical analysis concerning the non-standard BPS\ equations previously
presented by means of the relaxation technique. Also, we depict the
corresponding self-dual k-solutions for the electromagnetic sector and
comment on the main features they engender. Finally, in Sec.\ref{end}, we
present our ending comments and perspectives.

From now on, we use the natural units system and a plus-minus signature for
the planar Minkowski metric.


\section{The model}

\label{general}

{The planar Lagrangian density describing the generalized
Maxwell-Chern-Simons-Higgs model is given by}%
\begin{eqnarray}
\mathcal{L}_{d} &=&-\frac{h\left( \left\vert \phi \right\vert ,N\right) }{4}%
F_{\mu \nu }F^{\mu \nu }-\frac{k}{4}\epsilon ^{\mu \nu \rho }A_{\mu }F_{\nu
\rho }  \label{m} \\
&&\hspace{-1cm}+w\left( \left\vert \phi \right\vert \right) \left\vert
D_{\mu }\phi \right\vert ^{2}+\frac{h\left( \left\vert \phi \right\vert
,N\right) }{2}\partial _{\mu }N\partial ^{\mu }N-U\left( \left\vert \phi
\right\vert ,N\right) \text{,}  \notag
\end{eqnarray}%
{where }$F_{\mu \nu }=\partial _{\mu }A_{\nu }-\partial _{\nu }A_{\mu }${\
is the usual electromagnetic field strength tensor, }$D_{\mu }\phi =\partial
_{\mu }\phi +ieA_{\mu }\phi $ is the covariant derivative of the Higgs field
and $\epsilon ^{\mu \nu \rho }$ is the $\left( 1+2\right) -${dimensional
Levi-Civita's tensor (with }$\epsilon ^{012}=+1${)}. The additional real
scalar field $N$\ provides the stabilization of the self-dual solutions
arising in the presence of both Maxwell and Chern-Simons terms. {Here, }$%
h\left( \left\vert \phi \right\vert ,N\right) ${\ and }$w\left( \left\vert
\phi \right\vert \right) ${\ are positive-definite dimensionless functions
of the scalar fields of the model. The Higgs potential }$U\left( \left\vert
\phi \right\vert ,N\right) ${\ supporting spontaneous symmetry breaking is
supposed to have the following structure}%
\begin{equation}
U\left( \left\vert \phi \right\vert ,N\right) =\frac{k^{2}\left( N+G\left(
\left\vert \phi \right\vert \right) \right) ^{2}}{2h\left( \left\vert \phi
\right\vert ,N\right) }+e^{2}N^{2}\left\vert \phi \right\vert ^{2}w\left(
\left\vert \phi \right\vert \right) \text{,}  \label{p}
\end{equation}%
from which one notes that $U\left( \left\vert \phi \right\vert ,N\right) $
is defined in terms of $h\left( \left\vert \phi \right\vert ,N\right) $, $%
w\left( \left\vert \phi \right\vert \right) $ and $G\left( \left\vert \phi
\right\vert \right) $, with $\dim G=1/2$. It is worthwhile to point out that
the specific form of the potential in (\ref{p}) assures the self-duality of
the generalized model (\ref{m}), i.e., it is a consequence of the BPS
construction and can be derived from the energy density (\ref{xx}).

We introduce the mass scale $M$ of the model, and use it to perform the
scale transformations: $x^{\mu }\rightarrow M^{-1}x^{\mu }$, $\phi
\rightarrow M^{1/2}\phi $, $N\rightarrow M^{1/2}N$, $A^{\mu }\rightarrow
M^{1/2}A^{\mu }$, $k\rightarrow Mk$, $e\rightarrow M^{1/2}e$ and $\upsilon
\rightarrow M^{1/2}\upsilon $, where $\upsilon $ is the vacuum expectation
value of the Higgs field. Then, we get that $G\rightarrow M^{1/2}G$ and $%
\mathcal{L}_{d}\rightarrow M^{3}$$\mathcal{L}$ , and the modified model (\ref%
{m}) is now described by the dimensionless Lagrange density $\mathcal{L}$ ,
which\ has the same form as $\mathcal{L}_{d}$. Also, for simplicity, we
choose $e=\upsilon =k=1$.

It is well-known that Chern-Simons {\ theories only exhibit electrically
charged static solutions}. In fact, the static Gauss law related to (\ref{m}%
) is ($j$ runs over spatial indices only)%
\begin{equation}
\partial _{j}\left( h\partial ^{j}A^{0}\right) +2\left\vert \phi \right\vert
^{2}A^{0}w=F_{12}\text{,}  \label{gl}
\end{equation}%
from which one {\ observes} that, even in the presence of non-trivial $%
h\left( \left\vert \phi \right\vert ,N\right) $, $w\left( \left\vert \phi
\right\vert \right) $ and $G\left( \left\vert \phi \right\vert \right) $, {%
the absence of electrically uncharged static solutions still holds, since
the temporal gauge $\left( A^{0}=0\right) $ does not solve (\ref{gl}).}

{Now, since the gauge }$A^{0}=0$ {\ cannot be implemented and supposing that 
}$h\left( \left\vert \phi \right\vert ,N\right) $\textbf{, }$w\left(
\left\vert \phi \right\vert \right) ${\ and }$G\left( \left\vert \phi
\right\vert \right) ${\ are non-trivial functions,} {the looking for static
solutions for the} second-order Euler-Lagrange equations related to (\ref{m}%
) {can be a quite hard task}, even in the presence of suitable boundary
conditions. {\ In this context, the adequate implementation of the BPS
formalism is quite useful, since it helps to find finite energy
configurations as solutions of a given set of first-order differential
equations.}

So, from now on, we focus our attention on the development of a BPS
framework consistent with (\ref{m}) and (\ref{p}). Specifically, we look for
radially symmetric solutions using the standard static \textit{Ansatz}%
\begin{equation}
\phi \left( r,\theta \right) =g\left( r\right) e^{in\theta }\text{ \ and \ }%
N\left( r,\theta \right) =\mp A^{0}\left( r\right) \text{,}  \label{z1}
\end{equation}%
\begin{equation}
\mathbf{A}\left( r,\theta \right) =-\frac{\widehat{\theta }}{r}\left(
a\left( r\right) -n\right) \text{,}  \label{a2}
\end{equation}%
where $\left( r,\theta \right) $ are polar coordinates and $n=\pm 1$, $\pm 2$%
, $\pm 3$ ... is the winding number (vorticity) of the configuration. The
fields $g\left( r\right) $, $a\left( r\right) $ and $A^{0}\left( r\right) $
must obey the usual boundary conditions%
\begin{equation}
g\left( 0\right) =0\text{, \ \ }a\left( 0\right) =n\text{, \ \ }%
A_{0}^{\prime }\left( 0\right) =0\text{,}  \label{q1}
\end{equation}%
\begin{equation}
g\left( \infty \right) =1\text{, \ \ }a\left( \infty \right) =0\text{, \ \ }%
A_{0}\left( \infty \right) =0\text{,}  \label{q2}
\end{equation}%
where prime denotes the derivative with respect to $r$.

Now, we implement the Bogomol'nyi approach on the energy functional related
to (\ref{m}). Given the generalized potential (\ref{p}), the radially
symmetric non-standard energy density can be written as%
\begin{eqnarray}
\varepsilon &=&\frac{h}{2}\left( \frac{1}{r}\frac{da}{dr}\right)
^{2}+h\left( \frac{dA^{0}}{dr}\right) ^{2}+\frac{\left( G\mp A^{0}\right)
^{2}}{2h}  \notag \\
&&+w\left( \left( \frac{dg}{dr}\right) ^{2}+\frac{g^{2}a^{2}}{r^{2}}\right)
+2g^{2}A_{0}^{2}w\text{,}  \label{ge}
\end{eqnarray}%
{where }$h=h\left( g,A_{0}\right) $\textbf{, }$w=w\left( g\right) ${\ and }$%
G=G\left( g\right) ${. It is clear that, in order to guarantee a
positive-definite energy, both }$h${\ and }$w$ must be positive-definite and
finite. Now, by using the static Gauss law (\ref{gl}) together with the
constraint 
\begin{equation}
\frac{dG}{dg}=2wg\text{,}  \label{dc}
\end{equation}%
{the energy density (\ref{ge}) can be rewritten in the form}%
\begin{eqnarray}
\varepsilon &=&\frac{h}{2}\left( \frac{1}{r}\frac{da}{dr}\mp \frac{1}{h}%
\left( G\mp A^{0}\right) \right) ^{2}+w\left( \frac{dg}{dr}\mp \frac{ag}{r}%
\right) ^{2}  \notag \\
&&\pm \frac{1}{r}\frac{d}{dr}\left( aG\right) +\frac{1}{r}\frac{d}{dr}\left(
rhA^{0}\frac{dA^{0}}{dr}\right) \text{.}  \label{xx}
\end{eqnarray}

{At this point, we see that the resulting total energy is minimized by the
non-standard first-order equations}%
\begin{equation}
\frac{dg}{dr}=\pm \frac{ag}{r}\text{,}  \label{xx1}
\end{equation}%
\begin{equation}
\frac{1}{r}\frac{da}{dr}=\pm \frac{1}{h}\left( G\mp A^{0}\right) \text{.}
\label{xx2}
\end{equation}%
{Equations (\ref{xx1}) and (\ref{xx2}) are the BPS equations related to the
generalized MCS-Higgs model (\ref{m}).} In the present case, the BPS total
energy $E_{bps}$ related to the solutions of (\ref{xx1}) and (\ref{xx2}) can
be explicitly evaluated%
\begin{equation}
E_{bps}=2\pi \int r\varepsilon _{bps}dr=\pm 2\pi n\text{,}  \label{xx3}
\end{equation}%
{where the minimal energy density }$\varepsilon _{bps}$ {is defined as}%
\begin{equation}
\varepsilon _{bps}=\pm \frac{1}{r}\frac{d}{dr}\left( aG\right) +\frac{1}{r}%
\frac{d}{dr}\left( rhA^{0}\frac{dA^{0}}{dr}\right) \text{.}  \label{xx4}
\end{equation}%
{Also, to compute }the BPS energy (\ref{xx3}), {we have used the boundary
conditions (\ref{q1}) and (\ref{q2}) together with}%
\begin{equation}
G\left( r=0\right) =-1\text{, ~\ }G\left( r=\infty \right) =0\text{.}
\label{ggg}
\end{equation}%
Finally, we assume that $h\left( \left\vert \phi \right\vert ,N\right) $ and 
$w\left( \left\vert \phi \right\vert \right) $ behave as%
\begin{equation}
h\left( r=0\right) =H_{0}\text{, \ \ }h\left( r=\infty \right) =H_{\infty }%
\text{,}
\end{equation}%
\begin{equation}
w\left( r=0\right) =W_{0}\text{, \ \ }w\left( r=\infty \right) =W_{\infty }%
\text{,}
\end{equation}%
where $H_{0}$, $W_{0}$ and $H_{\infty }$ are real non-negative constants,
and $W_{\infty }$ is a real positive constant.

{From (\ref{xx3}), one notes that the total energy $E_{bps}$ is quantized
according the winding number $n$. Furthermore, it can be related to the
magnetic flux $\Phi _{B}$ in the standard way. Also, we point out that the
first-order framework developed here is implemented for any function $%
h\left( \left\vert \phi \right\vert ,N\right) $ finite and
positive-definite, i.e., given some function $h\left( \left\vert \phi
\right\vert ,N\right) $, the set formed by (\ref{xx1}) and (\ref{xx2}) gives
the BPS configurations of the generalized MCS-Higgs model (\ref{m}).} On the
other hand, in the absence of (\ref{p}) or (\ref{dc}), the energy functional
(\ref{ge}) cannot be written as (\ref{xx}), and the development of a
consistent non-standard first-order formalism cannot be performed.

{The prescription for the development of generalized self-dual
Maxwell-Chern-Simons-Higgs models is as follows: given any function }$%
w\left( \left\vert \phi \right\vert \right) ${, positive and finite, the
corresponding }$G\left( \left\vert \phi \right\vert \right) ${\ is obtained
by means of Eq. (\ref{dc}). Then, given any function }$h\left( \left\vert
\phi \right\vert ,N\right) ${, also positive and finite, the resulting
self-dual potential }$U(\left\vert \phi \right\vert ,N)${\ is determined via
(\ref{p}). In this scenario, the BPS states of the generalized model (\ref{m}%
) are the solutions of the equations (\ref{gl}), (\ref{xx1}) and (\ref{xx2}%
). The total energy is given by (\ref{xx3}), and the energy density by (\ref%
{xx4}).}

{In the standard MCS-Higgs case, BPS solutions with non-trivial topology
only exist in the asymmetric vacuum of the potential}%
\begin{equation}
U_{s}\left( \left\vert \phi \right\vert ,N\right) =\frac{1}{2}\left(
\left\vert \phi \right\vert ^{2}+N-1\right) ^{2}+N^{2}\left\vert \phi
\right\vert ^{2}\text{,}  \label{j4}
\end{equation}%
{which is defined by }$N=0${\ and }$\left\vert \phi \right\vert =1${. So,
for simplicity, we assume that also the generalized potential (\ref{p})
achieves its asymmetric vacuum when }$N=0${\ and }$\left\vert \phi
\right\vert =1$. As a consequence, topologically non-trivial self-dual
solutions of (\ref{m}) exist in the same asymmetric phase as their standard
counterparts.

{The standard MCS-Higgs model is trivially recovered starting from our
generalized framework by setting }$w\left( \left\vert \phi \right\vert
\right) =h\left( \left\vert \phi \right\vert ,N\right) =1$. So, instead of
recovering the usual model, we introduce a generalization of such theory.
The generalized model is defined by%
\begin{equation}
w\left( \left\vert \phi \right\vert \right) =b\left( \left\vert \phi
\right\vert ^{2}-1\right) ^{b-1}\text{,}  \label{j1}
\end{equation}%
where $b$ is a positive odd- number. The corresponding $G\left( \left\vert
\phi \right\vert \right) $ is%
\begin{equation}
G\left( \left\vert \phi \right\vert \right) =\left( \left\vert \phi
\right\vert ^{2}-1\right) ^{b}\text{.}  \label{j2}
\end{equation}%
For simplicity, we choose $h\left( \left\vert \phi \right\vert ,N\right) $
as follows%
\begin{equation}
h\left( N\right) =\alpha N^{2}+1\text{,}  \label{j3}
\end{equation}%
where $\alpha $ is a real non-negative number. Note that $\alpha =0$ and $b=1
$ leads us back to the standard case. The resulting generalized self-dual
potential $U\left( \left\vert \phi \right\vert ,N\right) $ (\ref{p}) is%
\begin{eqnarray}
U\left( \left\vert \phi \right\vert ,N\right)  &=&\frac{\left( N+\left(
\left\vert \phi \right\vert ^{2}-1\right) ^{b}\right) ^{2}}{2\left( \alpha
N^{2}+1\right) }  \notag \\
&&+bN^{2}\left\vert \phi \right\vert ^{2}\left( \left\vert \phi \right\vert
^{2}-1\right) ^{b-1}\text{.}
\end{eqnarray}%
{In the present case, (\ref{gl}), (\ref{xx1}) and (\ref{xx2}) can be
written, respectively, as}%
\begin{eqnarray}
&&\left( \alpha A_{0}^{2}+1\right) \left( \frac{d^{2}A^{0}}{dr^{2}}+\frac{1}{%
r}\frac{dA^{0}}{dr}\right) +2\alpha A^{0}\left( \frac{dA^{0}}{dr}\right) ^{2}
\notag \\
&&\text{ \ \ \ \ \ \ \ \ \ }\left. =2bg^{2}\left( g^{2}-1\right) ^{b-1}A^{0}-%
\frac{1}{r}\frac{da}{dr}\right. \text{,}  \label{xx5}
\end{eqnarray}%
\begin{equation}
\frac{dg}{dr}=\pm \frac{ag}{r}\text{,}  \label{j6}
\end{equation}%
\begin{equation}
\frac{1}{r}\frac{da}{dr}=\frac{\pm \left( g^{2}-1\right) ^{b}-A^{0}}{\alpha
A_{0}^{2}+1}\text{,}  \label{j7}
\end{equation}%
which must be solved according the finite energy boundary conditions %
\eqref{q1} and (\ref{q2}).

In the next Section, we solve \eqref{xx5}, \eqref{j6} and \eqref{j7} via %
\eqref{q1} and (\ref{q2}) for different values of $\alpha $ and $b$. Then,
we use the numerical solutions we found for $g\left( r\right) $, $a\left(
r\right) $ and $A^{0}\left( r\right) $ to depict the corresponding profiles
for the electric field%
\begin{equation}
E\left( r\right) =-\frac{dA^{0}}{dr}\text{,}  \label{ef}
\end{equation}%
the magnetic field%
\begin{equation}
B\left( r\right) =-\frac{1}{r}\frac{da}{dr}\text{,}  \label{mg}
\end{equation}%
and the minimal energy density (\ref{xx4}). Also, we comment on the main
features the non-standard numerical solutions engender.


\section{Numerical solutions}

\label{numerical2}

In this Section, we focus our attention on the non-standard numerical
solutions themselves. The equations to be studied are \eqref{xx5}, \eqref{j6}
and \eqref{j7}, and the fields $g\left( r\right) $, $a\left( r\right) $ and $%
A^{0}\left( r\right) $ must behave according the finite energy boundary
conditions \eqref{q1} and (\ref{q2}).

Here, for simplicity, we choose $n=1$. {Then, we solve the first-order
system numerically by means of the relaxation technique, for different
values of the real parameters }$\alpha ${\ and }$b${. The profiles of the
physical fields and energy density\ are depicted in the figures below.} The
system was solved for $\alpha =0$ and $b=1$ (usual case, dash-dotted black
line), $\alpha =0$ and $b=3$ (dotted blue line), $\alpha =5$ and $b=1$
(dashed red line) and $\alpha =5$ and $b=3$ (solid green line). 
\begin{figure}[tbph]
\centering\includegraphics[width=8.5cm]{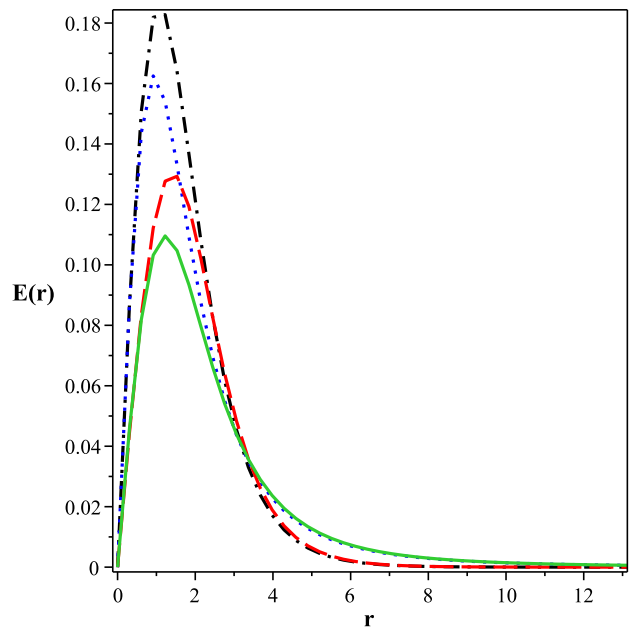}
\par
\vspace{-0.3cm}
\caption{Solutions to $E\left( r\right) $ for $\protect\alpha =0$ and $b=1$
(dash-dotted black line), $\protect\alpha =0$ and $b=3$ (dotted blue line), $%
\protect\alpha =5$ and $b=1$ (dashed red line) and $\protect\alpha =5$ and $%
b=3$ (solid green line).}
\end{figure}

In Figure 1, we depict the numerical solutions for the electric field %
\eqref{ef}, and we see that the non-standard ones behave, in general, as
their usual counterpart: starting from $0$ (zero), the solutions reach their
maximum values at some finite distance $R$ from the origin, and vanish as $r$
goes to infinity. However, we point out that different solutions exhibit not
only different amplitudes, but also slightly different characteristic
lengths: {in general, increments on }$\alpha ${\ and/or }$b${\ lead to
decrements on amplitudes and/or characteristic length}. Even in this case,
such variations are expected to occur in the context of a non-standard
theory as (\ref{m}); see, for instance, \cite{n14b}. 
\begin{figure}[tbph]
\centering\includegraphics[width=8.5cm]{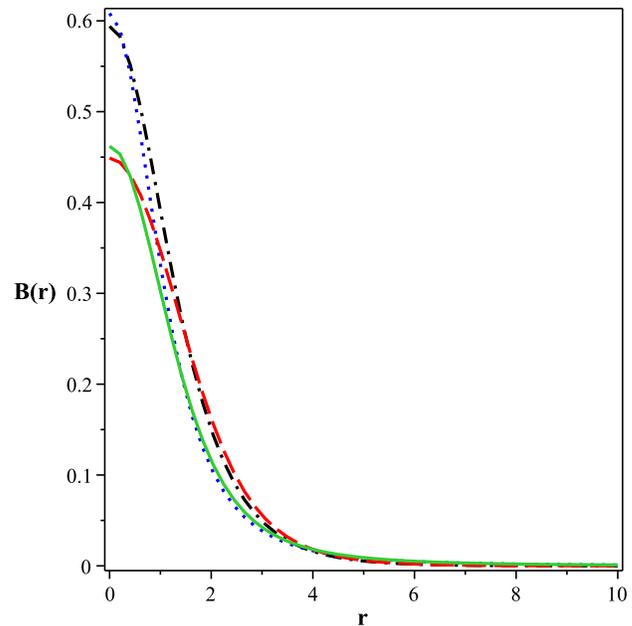}
\par
\vspace{-0.3cm}
\caption{Solutions to $B\left( r\right) $. Conventions as in FIG. 1.}
\end{figure}

In Figure 2, we present the profile of the magnetic field (\ref{mg}). The
same manner as for the electric field, the generalized solutions behave as
the standard one: the magnetic fields are lumps centered at the origin, and
they decrease monotonically as $r$ goes to infinity.{\ In this case, for a
fixed }$\alpha ${\ and an increasing }$b${, the amplitudes experience a
small increase and the characteristic length decreases. On the other hand,
for a fixed }$b${\ and\ an increasing }$\alpha ${, the amplitude will
decrease significantly, but the characteristic lengths remain approximately
fixed by a compensatory effect relating }$\alpha ${\ and }$b${.} 
\begin{figure}[tbph]
\centering\includegraphics[width=8.5cm]{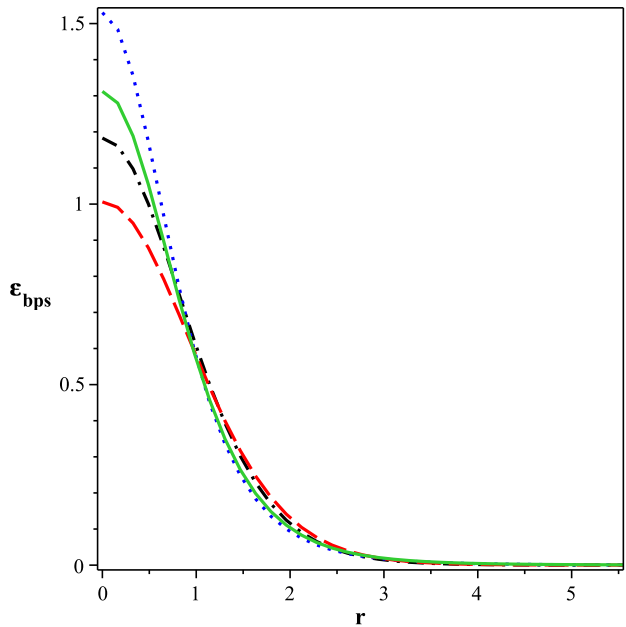}
\par
\vspace{-0.3cm}
\caption{Solutions to $\protect\varepsilon _{bps}$. Conventions as in FIG.
1. }
\end{figure}

Numerical solutions for the minimal energy density $\varepsilon _{bps}$ (\ref%
{xx4}) are {depicted} in Figure 3, and we see that also such solutions are
lumps centered at $r=0$ which decrease monotonically to $r\rightarrow 0$. {%
The lumps profile are similar to those the magnetic field. However, in this
case, for a fixed $\alpha $ and an increasing $b$, the amplitudes experience
a large increase and, the characteristic lengths decreased slightly. On the
other hand, for a fixed $b$ and an increasing $\alpha $, the amplitude will
decrease significantly but the characteristic lengths remain almost fixed by
a compensatory effect relating $\alpha $ and $b$.} 
\begin{figure}[tbph]
\centering\includegraphics[width=8.5cm]{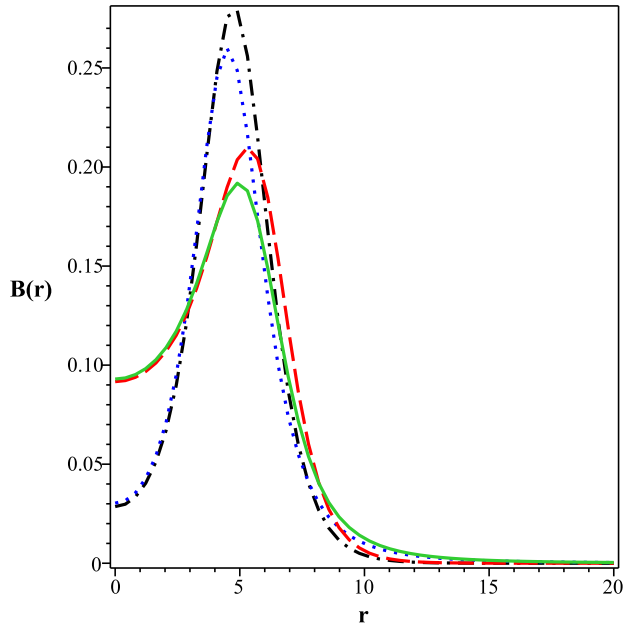}
\par
\vspace{-0.3cm}
\caption{Solutions to $B\left( r\right) $ for $n=5$. Conventions as in FIG.
1.}
\end{figure}

{To end this Section, we have depicted the numerical profiles for }$B\left(
r\right) ${\ and }$\varepsilon _{bps}${\ for }$n=5${\ (see figures 4 and 5,
respectively). It is well known that in the standard model, for a large
winding number, the profiles of both the magnetic field and the energy
density go from a lump (}$n=1${) to a ring (}$n>>1$\textbf{)}. From figures
4 and 5, one notes that also in this limit (increasing vorticity), the
generalized solutions mimic the usual ones, as expected. For large $n$\ {%
with fixed $\alpha $ and a increasing $b$, the profiles experience a slight
change. However for $b$ fixed and an increment in $\alpha $, the profile's
amplitudes experience a considerable decrease.}


\section{Ending comments}

\label{end}

In the present letter, we have performed the development of a first-order
theoretical framework in the context of a generalized MCS-Higgs model given
by (\ref{m}). In the present case, the non-standard model, including its
self-dual potential (\ref{p}), is given in terms of three different
functions, $h\left( \left\vert \phi \right\vert ,N\right) $, $w\left(
\left\vert \phi \right\vert \right) $ and $G\left( \left\vert \phi
\right\vert \right) $, which are functions of the scalar fields only. Here,
in order to avoid problems with the energy of the model, $h\left( \left\vert
\phi \right\vert ,N\right) $ and $w\left( \left\vert \phi \right\vert
\right) $ must be {positive-definite}. Then, given the general structure for
the self-dual potential (\ref{p}), the consistence of the non-standard
first-order approach only holds when $w\left( \left\vert \phi \right\vert
\right) $ and $G\left( \left\vert \phi \right\vert \right) $ are {related by
the differential} constraint (\ref{dc}). On the other hand, one notes that {%
there is no additional }constraint to be imposed on $h\left( \left\vert \phi
\right\vert ,N\right) $. 
\begin{figure}[tbph]
\centering\includegraphics[width=8.5cm]{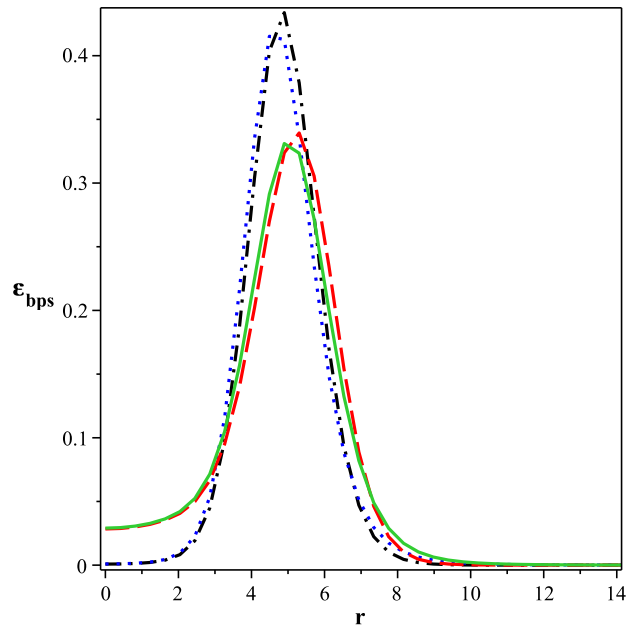}
\par
\vspace{-0.3cm}
\caption{Solutions to $\protect\varepsilon _{bps}$ for $n=5$. Conventions as
in FIG. 1.}
\end{figure}

{After perform the construction of the generalized MCS-Higss model, we
illustrate such realization providing an explicit non-standard model
specified by (\ref{j1}), (\ref{j2}) and (\ref{j3}). In this case, the
non-standard model is controlled by two real parameters, $\alpha $ and $b$.
Immediately, we have investigated radially symmetric self-dual
configurations given by the usual static \textit{Ansatz} (\ref{z1}) and (\ref%
{a2}), and considering that the fields behave according the usual finite
energy boundary conditions (\ref{q1}) and (\ref{q2}).}

We have integrated the generalized BPS equations by means of the relaxation
technique, for different values of $\alpha $ and $b$. The numerical profiles
we found are depicted in figures 1, 2, 3, 4 and 5. In general, we have seen
that the non-standard solutions mimic their usual counterparts. Also, as
expected, we have noted variations on the amplitudes and on the
characteristic lengths {of the modified solutions}. {Furthermore, we have
identified the rules controlling such variations.}

This work is an extension of a recent investigation performed by some of us 
\cite{n14d}. In this sense, a very interesting issue to be considered
concerns the development of a non-standard self-dual MCS-Higgs model which
exhibits the very same numerical solutions engendered by the standard
theory. In this case, such investigation would be a generalization of the
one presented in \cite{n14e}, regarding a non-standard Maxwell-Higgs model.
Another issue concerns the supersymmetric extension of the self-dual model (%
\ref{m})-(\ref{p}).\ {These issues are in advance, }and we hope report
interesting results in a near future.

The authors would like to thank CAPES, CNPq and FAPEMA (Brazil) and FCT
Project CERN/FP/116358/2010 (Portugal) for partial financial support. Also,
we are grateful to C. dos Santos for useful discussions. E. da Hora thanks
the Department of Mathematical Sciences of Durham University (U.K.), for all
their hospitality while doing part of this work.

\end{document}